\documentstyle[12pt]{article}

\title{
\begin{flushright}
{\normalsize Yaroslavl State University\\
             Preprint YARU-HE-99/03\\
             hep-ph/9904453} \\[5mm]
\end{flushright}
Neutrino - electron processes in a strong magnetic field 
and plasma}

\author{{A.V.~Kuznetsov, N.V.~Mikheev}\\ [7mm] 
{\small\it Division of Theoretical Physics, Department of Physics,}\\
{\small\it Yaroslavl State University, Sovietskaya 14,}\\
{\small\it 150000 Yaroslavl, Russian Federation}}

\date{}

\begin{document}

\maketitle

\begin{abstract} 
The total set of the neutrino - electron processes 
in a presence of both components of an active medium, hot dense plasma 
and strong magnetic field, is investigated for the first time. 
The contribution of the processes $\nu e^- \to \nu e^-$, 
$\nu e^+ \to \nu e^+$, $\nu \to \nu e^- e^+$, $\nu e^- e^+ \to \nu$ is 
shown to dominate and not to depend on the chemical potential of 
electron-positron gas. Relatively simple expressions for the probability 
and mean losses of the neutrino energy and momentum are obtained, 
which are suitable for a quantitative analysis. 
\end{abstract}

\newpage

An understanding of the important role of neutrino interactions in 
astrophysical processes stimulates a constantly growing interest in the 
neutrino physics in a dense medium~\cite{Raff}. 
For example, in a collapsing stellar core where a large number of neutrinos 
is produced, the density amounts up to the value of a nuclear density, and 
matter becomes opaque to neutrinos. The elastic scattering on nuclei was 
usually considered as the main source of neutrino opacity. A contribution 
of the neutrino-electron scattering $\nu e^- \to \nu e^-$ was estimated to 
be essentially smaller, and it was not included in the earlier attempts of 
collapse simulation (see, e.g.,~\cite{Coop} and references therein). 
However, as the analysis showed~\cite{Myra}, neu\-tri\-no-elec\-tron 
scattering can contribute 
essentially to the energy balance of the collapsing stellar core. 
A numerical computation of the differential probability of the 
reaction $\nu e^- \to \nu e^-$ 
in degenerate electron plasma was performed in a paper~\cite{Bez}, 
where an influence of relatively small magnetic fields was taken into account.

It should be noted that a consideration of a strong magnetic field
as a medium, along with 
a dense matter, is physically justified indeed. Really, the field 
strengths inside the astrophysical objects can reach the critical 
Schwinger value $B_e = m_e^2/e \simeq 4.41 \cdot 10^{13}$ G
\footnote{We use natural units in which $c = \hbar = 1.$}
, and even exceed it essentially. 
In the present view, very strong magnetic fields at a level of
$10^{16} $G can exist inside the astrophysical 
cataclysms like a supernova explosion or a coalescence of neutron stars. 
For example, such fields could be generated in a 
supernova envelope due to the mechanism suggested by 
Bisnovatyi-Kogan~\cite{Bis}. 
It should be emphasized that such field is really rather dense medium with 
the mass density 
\begin{equation}
\rho \, = \, \frac{B^2}{8 \pi} \, \simeq \, 0.4 \cdot 10^{10} 
{\mbox{g} \over \mbox{cm}^3} \cdot 
\left({B \over {10^{16} \mbox{G}}}\right)^2 ,
\label{eq:rho}
\end{equation}

\noindent 
which is comparable with the plasma mass density $10^{10} - 10^{12} 
\mbox{g/cm}^3$ to be typical for the envelope of an exploding supernova. 
It is known that such intense fields make an active influence 
on quantum processes, thus allowing the transitions which are kinematically 
forbidden in vacuum. 

In this paper we investigate the total set of
the neutrino - electron processes 
in a presence of both components of an active medium, hot dense plasma 
and strong magnetic field. This set includes not only the ``canonical'' 
scattering $\nu e^- \to \nu e^-$, $\nu e^+ \to \nu e^+$, 
$e^+ e^- \leftrightarrow \nu \bar\nu$, but the ``exotic'' process of 
the synchrotron emission of the neutrino pair and the reverse process, 
$e \leftrightarrow e \nu \bar\nu$, and the processes of 
production and absorption of the elec\-tron-po\-si\-tron pair 
$\nu \to \nu e^- e^+$, $\nu e^- e^+ \to \nu$ as well. 
The ``exotic'' processes become opened only in the presence of a magnetic 
field. 

We consider the physical situation when the field strength $B$ appears to be 
the largest physical parameter, while the electron mass is the smallest one
\begin{equation}
e B \, \gg \, E^2, \, \mu^2, \, T^2 \, \gg \, m_e^2,
\label{eq:eB}
\end{equation}

\noindent where $E$ is the neutrino energy, 
$\mu$ is the chemical potential of electrons,
$T$ is the temperature of plasma. In this limit, the electrons and the 
positrons occupy the lowest Landau level. 

We use the effective Lagrangian of the neutrino - electron interaction 
in the local limit 

\begin{equation}
{\cal L} \, = \, \frac{G_F}{\sqrt 2} 
\big [ \bar e \gamma_{\alpha} (g_V - g_A \gamma_5) e \big ] \,
\big [ \bar \nu \gamma^{\alpha} (1 - \gamma_5) \nu \big ] \,,
\label{eq:L}
\end{equation}

\noindent where $g_V = \pm 1/2 + 2 sin^2 \theta_W, \, g_A = \pm 1/2$.
Here the upper signs correspond to the electron neutrino 
($\nu = \nu_e$) when both $Z$ and $W$ boson exchange takes part 
in a process. The lower signs correspond to $\mu$ and $\tau$ neutrinos 
($\nu = \nu_{\mu}, \nu_{\tau}$), when the $Z$ boson exchange 
is only presented in the Lagrangian~(\ref{eq:L}). 

An amplitude of any above-mentioned 
neutrino - electron process could be immediately 
obtained from the Lagrangian~(\ref{eq:L}) where the known solutions of the 
Dirac equation in a magnetic field should be used. 
At first glance all the neutrino - electron processes should be strongly 
suppressed in the considered limit~(\ref{eq:eB}), because the amplitude 
tends to zero in the limit of massless electron. It can be best demonstrated 
for the processes $e^+ e^- \leftrightarrow \nu \bar\nu$. 
Really, the total spin of the neutrino - antineutrino 
pair in the center-of-mass system is equal to 1, while 
the total spin of the electron - positron pair on the lowest Landau level 
is equal to zero. Thus the process amplitude is strictly zero for the case of 
the local interaction and massless particles, and it is suppressed 
in the considered relativistic limit. 

However, as the analysis shows, all the neutrino - electron processes 
described by the Lagrangian~(\ref{eq:L}) can be separated into two parts:

i) the processes with the neutrino - antineutrino pair in the initial or 
in the final state, $e^- e^+ \leftrightarrow \nu \bar\nu$, 
$e \leftrightarrow e \nu \bar\nu$, where the above-mentioned suppression 
remains valid after integration over the phase space;

ii) the processes where the neutrino presents both in the initial and 
in the final state, $\nu e^\mp \to \nu e^\mp$, 
$\nu \leftrightarrow \nu e^- e^+$, and the similar processes with the 
antineutrino, where a non-trivial kinematic enhancement takes place in 
the integration over the phase space, which provides the exact compensation   
of the amplitude suppression. 

We investigate first the ``canonical'' process of neutrino scattering 
on the electrons of magnetized plasma. 
The probability of this process in a unit time has a physical 
meaning only being integrated over both the final and the initial electron 
states as well
\begin{equation}
W (\nu e^- \to \nu e^-) = \frac{1}{\cal T} \int |{\cal S}|^2 
\; d \Gamma_{e^-} \; f_{e^-} 
\; d \Gamma'_{e^-} \; (1 - f'_{e^-}) 
\; d \Gamma'_\nu \; (1 - f'_\nu), 
\label{eq:Wdef}
\end{equation}

\noindent 
where $\cal T$ is the total interaction time, 
$\cal S$ is the matrix element of the transition, 
$d \Gamma$ is the phase-space element of a particle, 
$f$ is its distribution function, 
$f'_\nu = [exp((E'-\mu_\nu)/T_\nu) + 1]^{-1}$, 
$\mu_\nu, T_\nu$ are the chemical potential and the temperature of 
the neutrino gas, 
$f_{e^{\mp}} = [exp((E_{\mp} \mp \mu)/T) + 1]^{-1}$, 
$\mu, T$ are the chemical potential and the temperature of the 
elec\-tron-po\-si\-tron gas. 
We do not present here the details of integration over the phase space of 
particles, which will be published in an extended paper. The result of our 
calculation of the probability~(\ref{eq:Wdef}) can be presented in the form 
\begin{eqnarray}
&& W (\nu e^- \to \nu e^-) = 
\frac{G_F^2 e B T^2 E}{4 \pi^3} 
\nonumber\\
&&\times \bigg \lbrace 
(g_V + g_A)^2 (1 - u)^2 
\int\limits_0^{x \tau \frac{1 + u}{2}} 
\frac{d \xi}{(1 - e^{-\xi}) (1 + e^{-x + \eta_\nu + \xi/\tau})} \,
\ln{\frac{1 + e^\eta}{1 + e^{-\xi+\eta}}}
\nonumber\\
&&+ 
(g_V - g_A)^2 (1 + u)^2 
\int\limits_0^{x \tau \frac{1 - u}{2}} 
\frac{d \xi}{(1 - e^{-\xi}) (1 + e^{-x + \eta_\nu + \xi/\tau})} \,
\ln{\frac{1 + e^\eta}{1 + e^{-\xi+\eta}}}
\nonumber\\
&& + 
[ (g_V + g_A)^2 (1 - u)^2 + (g_V - g_A)^2 (1 + u)^2]
\nonumber\\
&&\times 
\int\limits_0^\infty 
\frac{d \xi}{(e^\xi - 1) (1 + e^{-x + \eta_\nu - \xi/\tau})}\,
\ln{\frac{1 + e^\eta}{1 + e^{-\xi+\eta}}}
 \bigg \rbrace ,
\label{eq:Wsc}
\end{eqnarray}

\noindent 
where $x = E/T_\nu$, $\tau = T_\nu/T$, 
$\eta_\nu = \mu_\nu/T_\nu$, $\eta = \mu/T$, $u = \cos \theta$, 
$\theta$ is an angle between the initial neutrino momentum
$\bf p$ and the magnetic field induction $\bf B$. 
The dependence of the scattering probability~(\ref{eq:Wsc}) 
on the electron density, $n = n_{e^-} - n_{e^+}$, is defined by the 
dependence on the chemical potential of electrons
$$\mu = \frac{2 \pi^2 n}{e B}.$$ 
Let us note that the expression~(\ref{eq:Wsc}) is valid 
for the cases of both hot ($\mu \ll T$) and cold ($\mu \gg T$) plasma.
The probability of the neutrino scattering on positrons, 
$\nu e^+ \to \nu e^+$, 
can be obtained from Eq.~(\ref{eq:Wsc}) by a simple change of sign of 
the electron chemical potential, $\mu \to - \mu$.

The probabilities of the remaining ``exotic'' processes 
$\nu \leftrightarrow \nu e^- e^+$ are defined similarly to~(\ref{eq:Wdef}) 
with the substitution $f_{e^-} \to (1 - f_{e^+})$ in every transposition 
of the electron from the initial (final) state to the final (initial) one. 
The results of our calculations are
\begin{eqnarray}
&& W (\nu \to \nu e^- e^+) =
\frac{G_F^2 e B T^2 E}{4 \pi^3} 
\label{eq:Wcr}\\
&&\times \bigg \lbrace 
(g_V + g_A)^2 (1 - u)^2 
\int\limits_0^{x \tau \frac{1 + u}{2}} 
\frac{d \xi}{(1 - e^{-\xi}) (1 + e^{-x + \eta_\nu + \xi/\tau})} \,
\ln{\frac{\cosh{\xi} + \cosh{\eta}}{1 + \cosh{\eta}}}
\nonumber\\
&&+ 
(g_V - g_A)^2 (1 + u)^2 
\int\limits_0^{x \tau \frac{1 - u}{2}} 
\frac{d \xi}{(1 - e^{-\xi}) (1 + e^{-x + \eta_\nu + \xi/\tau})} \,
\ln{\frac{\cosh{\xi} + \cosh{\eta}}{1 + \cosh{\eta}}}
 \bigg \rbrace ,
\nonumber
\end{eqnarray}

\vspace{2mm}

\begin{eqnarray}
&& W (\nu e^- e^+ \to \nu) =
\frac{G_F^2 e B T^2 E}{4 \pi^3} 
[ (g_V + g_A)^2 (1 - u)^2 + (g_V - g_A)^2 (1 + u)^2]
\nonumber\\
&&\times \int\limits_0^\infty 
\frac{d \xi}{(e^\xi - 1) (1 + e^{-x + \eta_\nu - \xi/\tau})} 
\, \ln{\frac{\cosh{\xi} + \cosh{\eta}}{1 + \cosh{\eta}}}.
\label{eq:Wab}
\end{eqnarray}

We note that rather the total probability of the neutrino interaction 
with magnetized elec\-tron-po\-si\-tron fraction of plasma

\begin{eqnarray}
W (\nu \to \nu) & = &
W (\nu \to \nu e^- e^+) + W(\nu e^- e^+ \to \nu) +
\nonumber\\
& + & W (\nu e^- \to \nu e^-) + W (\nu e^+ \to \nu e^+) 
\label{eq:Wtotd}
\end{eqnarray}

\noindent 
has a physical meaning, than the probabilities~(\ref{eq:Wsc}), (\ref{eq:Wcr}), 
(\ref{eq:Wab}) separately. For the total probability we obtain 
a more simple expression

\begin{eqnarray}
W (\nu \to \nu)  &=& 
\frac{G_F^2 e B T^2 E}{4 \pi^3} 
\bigg \lbrace 
(g_V + g_A)^2 (1 - u)^2 
\left[ F_1 \left(\frac{x \tau (1 + u)}{2} \right) - F_1 (- \infty) \right]
\nonumber\\
&+&
\left(g_A \to - g_A; \; u \to - u \right) 
 \bigg \rbrace ,
\label{eq:Wtot}
\end{eqnarray}

\noindent 
where
\begin{eqnarray}
F_k (z) = \int\limits_0^z
\frac{\xi^k d \xi}{(1 - e^{-\xi}) (1 + e^{-x + \eta_\nu + \xi/\tau})}. 
\label{eq:F(z)}
\end{eqnarray}

\noindent 
Both the probability~(\ref{eq:Wcr}) and
(\ref{eq:Wtot}) do reproduce our result~\cite{KM97a,KM97b} obtained 
for the case of a pure magnetic field, in the limit of a rarefied plasma 

\noindent 
($T,\,\mu,\, T_\nu,\, \mu_\nu \to 0$). 

It is interesting to note that the dependence on the electron chemical 
potential magically cancelled in the total probability~(\ref{eq:Wtot}), 
whereas each of the partial probabilities~(\ref{eq:Wsc}), (\ref{eq:Wcr}), 
(\ref{eq:Wab}) does depend on $\mu$. 
It means that the total probability does not depend on the electron 
density $n$ in the strong field limit~(\ref{eq:eB}). 
We do not know a physical underlying reason of this independence up to now. 
Probably, some property of a completeness of the considered set of processes 
with respect to the electrons manifests itself here. 

The probability~(\ref{eq:Wtot}) defines the partial contribution of the 
considered processes into the neutrino opacity of the medium. 
The estimation of the neutrino mean free path with respect to the 
neu\-tri\-no-elec\-tron processes yields
\begin{equation}
\lambda_e = {1 \over W} \simeq 170 \;\mbox{km} \cdot
\left ({{10^3 B_e} \over B} \right ) \;
\left ({{5\;\mbox{MeV} \over T}} \right )^3.
\label{eq:lambda}
\end{equation}

\noindent 
It should be compared with the mean free path caused by the interaction 
with nuclei, which is evaluated to be of order of 1 km at the density 
value $\rho \sim 10^{12} \;\mbox{g/cm}^3$. At first glance the influence 
of the neu\-tri\-no-elec\-tron reactions on the process of neutrino 
propagation is negligibly 
small. However, a mean free path does not exhaust the neutrino physics in 
a medium. The mean values of the neutrino energy and momentum loss are also 
essential in astrophysical applications. 

The mean values of the neutrino energy and momentum loss
\footnote{In general, a neutrino can both loose and absorb the energy and 
momentum.}
could be defined by the four-vector 
\begin{equation}
Q^{\alpha} \, = - \, 
E \left (\frac{d E}{d t}, \frac{d {\bf p}}{d t} \right )
 \, = \, E \int q^\alpha d W, 
\label{eq:Qal}
\end{equation}

\noindent where $E$ and $\bf p$ are the neutrino energy and momentum, 
$q$ is the difference of the momenta of the initial and final neutrinos, 
$q = p - p'$, $d W$ is the total differential probability of the processes 
considered. 
The zero component $Q^0$ is connected with the mean neutrino energy 
loss in a unit time, the space components $\bf Q$ are connected similarly 
with the neutrino momentum loss in a unit time. 

The four-vector $Q^{\alpha}$ was calculated in our papers~\cite{KM97a,KM97b}
for the case of a pure magnetic field. The losses are connected in this case 
with the pair production by a neutrino propagating in a strong magnetic field, 
$\nu \to \nu e^- e^+$, which is the only possible process in the absence 
of plasma. 

The result of our calculation of the zeroth and third components 
(the magnetic field is directed along the third axis) 
of the four-vector $Q^{\alpha}$ in a strongly magnetized plasma is
\begin{eqnarray}
Q_{0,3}  &=&  
\frac{G_F^2 e B T^3 E^2}{4 \pi^3} \;
\bigg \lbrace 
(g_V + g_A)^2 (1 - u)^2 
\left[ F_2 \left(\frac{x \tau (1 + u)}{2} \right) - F_2 (- \infty) \right]
\nonumber\\
&\pm&
\left(g_A \to - g_A; \; u \to - u \right) 
 \bigg \rbrace ,
\label{eq:Q03p}
\end{eqnarray}

\noindent 
where the function $F_2 (z)$ is defined in Eq.~(\ref{eq:F(z)}), 
the plus or minus signs correspond to the zeroth and third components. 
Our result for the four-vector of losses obtained in 
the case of a pure magnetic field~\cite{KM97a,KM97b}, is reproduced from 
Eq.~(\ref{eq:Q03p}) in the limit of a rarefied 
plasma ($T,\, T_\nu,\, \mu_\nu \to 0$). 

Finally, we have investigated the neutrino-electron interactions 
in strongly magnetized plasma, taking into account the 
total set of neutrino - electron 
processes, both ``canonical'', $\nu e^\mp \to \nu e^\mp$, 
and ``exotic'', $\nu \leftrightarrow \nu e^- e^+$.
The total probability and the four-vector of the mean values of the 
neutrino energy and momentum loss have been calculated. 
The surprising result is that 
these values appear not to depend on the electron density. 

There is good reason to believe that the results obtained could be useful 
in astrophysical applications. 

\bigskip

\noindent 
{\bf Acknowledgements}  

We are grateful to G.~Raffelt, V.~Rubakov, and V.~Semikoz for helpful 
discussions. 

This work was supported in part by the INTAS Grant N~96-0659  
and by the Russian Foundation for Basic Research Grant N~98-02-16694.
The work of A.K. was supported in part by the
International Soros Science Education Program under the Grant N~d99-76.

\end{document}